# An accurate detection of micro-collapse during the lyophilisation of a 5% w/v lactose solution using a combination of novel techniques : intelligent laser speckle imaging (ILSI) and through-vial impedance spectroscopy (TVIS)

Ahmet Orun[a], Anand Vadesa[b], Geoff Smith[b]*

[a]School of Computer Science and Informatics, Faculty of Computing Engineering and Media, De Montfort University, Leicester UK, LE1 9BH

[b]DMU LyoGroup, School of Pharmacy. Faculty of Health and life Sciences, De Montfort University, Leicester UK, LE1 9BH, E-mail address: gsmith02@dmu.ac.uk,  Phone: +44 (0)116 250 6298

## Abstract

Context: In a freeze drying *(FD)* process, an accurate observation and control of the process parameters at critical stages are at high importance. Particularly accurate and timely identification of the critical temperature ($T_c$) at the end of primary drying phase would lead to heating energy cost reduction and product waste elimination by ending the process before the product's micro-collapse stage. **Aim:** Within this work, a combination of novel techniques optical technique called Intelligent Laser Speckle Imaging *(ILSI)* in association with product image texture analysis, coupled to an electrical impedance technique called through vial impedance spectroscopy (TVIS) has been applied onto the real-time product states observation process to identify the specific structural product surface/subsurface characteristics and hence the micro-collapse stage. **Method:** 2 cycles – one providing a profile for standard approach (non-collapsed) and another with a temperature ramp through Tc to micro-collapse, TVIS provides an assessment of the onset of micro-collapse through the assessment of the acceleration in drying rate whereas the ILSI with pattern recognition detects the change in microstructure

## 1. Introduction

As is well known, in freeze drying *(FD)* process, an accurate parametric observation and control of each critical *FD* stage are very important factor for the product stabilization. Particularly accurate and timely tracking and detection of the critical temperature ($T_c$) at the end of primary drying stage would lead to heating energy cost reduction and final product waste elimination by immediate ending the process before the product micro-collapse stage. To achieve this, within this work a novel optical technique called Intelligent Laser Speckle Imaging (*ILSI*) [1][2][9] in association with image texture analysis has been applied to FD process stages of a product by direct visual observation and quantification rather than indirect sensor based (pressure, temperature, humidity, etc.) approach where some of them may be invasive or have low signal-to-noise ratio to contribute to the target model.

Many earlier studies had already focused on accurate determination of Critical Temperature ($T_c$) as it plays an important role in optimisation of the overall freeze drying process including the substantial reduction of primary drying period. Greco et al. [3] attempted to predict critical temperature accurately by use of Optical Coherence Tomography freeze-drying microscopy (OCT-FDM) which had already advantages over the conventional light FD microscopy. The study acknowledged a significant reduction of primary drying time and hence a production cost. However, the major disadvantage of the work was already reported  that microscopy scale experiments have substantial limitations of realistic or accurate parametric FD  assessments in comparison to vial scale experiments. Because the  freeze drying microscopy of 2D dimensional sample is different than FD sample in a vial (5°-10° C difference is reported for high protein formulation)[3]. Our FD experiments have been achieved in vials as is more similar to  actual industrial production environment. Meanwhile in another study conducted by Pikal and Shah [4] it was reported that Tc point for vial scale analysis is also higher than FD microscopy. Some of the studies have also proven that,  in primary drying process 1° C temperature increase corresponds to 10% reduction in primary drying time [3].  Different characteristics of collapse temperature for lactose product was also investigated in a Freeze-drying-microscopy related work [12] introduced by Raman. He stated that the collapse first occurs at the front of product as it corresponded to highest moisture content in lactose matrix which was observable by light band formation in freeze drying microscopy.

In the literature the majority of "FD-process-in-vial" studies introduce a parametric FD model (like LyoMonitor [13] in which many variables (e.g. pressure, temperature, moisture, etc.) are unified and exploited as they are often relying only on the sensory information that are indirectly related to the product states (e.g. restricted to FD chamber environment) rather than sensing the direct product characteristics such as its structural imaging. Whereas this would only occur by limited number of measurement methods like ultrasound [16], FD microscopy [3] with its unrealistic small scale process characteristics, X-ray analysis [17], etc. Some other methods in which FD product is directly contacted or mechanically treated (e.g. Texture analyser [14]) would not be practical for a real-time product observation during the entire FD process nor non-invasive. Our proposed non-invasive *ILSI* technique promises a close observation of direct product structure and characteristics to define an accurate micro-collapse state and $T_c$ for a substantial primary drying time reduction.

The methods and instrumental techniques introduced within this work propose a novel laser speckle related optical observation approach (image texture analysis in association with AI methods) to control lyophilisation process parameters particularly Critical Temperature ($T_c$) before the start of micro-collapse stage.

The proposed novel method (ILSI) is complementary one to the well-known "Through Vial Impedance Spectroscopy (TVIS)" method as ILSI is based on a direct observation of the product stages by a CCD camera, whereas TVIS is only relying on sensor measurements (e.g. impedance, temperature, etc.) which is an indirect method instead. In example, for TVIS method two different product states like micro-collapse and non-micro-collapse may yield a same sensor measurement value ($C''_{PEAK}$) as seen in Figure DSD, which leads to misidentification problem of a product state. For such circumstances ILSI can help specify the different product states at high accuracy (up to 95-100%) as a complementary method

## 2. Methods and Materials

### 2.1 Solution preparation

D-Lactose monohydrate was purchased from Fluka (UK) and used, as supplied, in the preparation of 5% w/v lactose solution in ultrapure water. 3.4 g aliquots of 5% w/v lactose solution to cluster of twenty-seven 10 ml type 1 tubular clear glass vial (Schott, Hungary), to two TVIS measurement vials and two cut 10 ml glass vials. Remaining glass vial filled with 3.4 g ultrapure water.

### 2.2 Freeze-drying

Two sorts of complementary FD experiments have been designed to gather various parametric system data along the process including TVIS and LSI methods whose main parameters are shown in Table 1 and
Table 2. We have demonstrated two FD experiments with sublimation one above and one below Tg' of lactose formulation. The first experiment (Figure1) targets a clear distinction between the two states of the FD product like non-collapse and micro-collapse forms. Where the corresponding FD time gaps are 17.5-19h and 20.5-22.5h respectively. The 2. Experiment (Figure 2) rather focuses on different forms of dry layer whose thickness varies in regards to FD parameters option and in this case the main goal is to test different dry layer thickness to develop a "dry layer volume invariant" method by avoiding any possibility of dry layer volumetric correlation effect on micro-collapse state identification by LSI method. By this approach, LSI method (or Intelligent LSI by the inclusion of AI classifiers) would be capable of detecting a micro collapse state of FD product against any volumetric effect (thickness) of dry layer state. In both experiments we have used two annealing steps so we can use 2. annealing data for temperature calibration. The temperature calibration data help to calculate $\hat{C}''$peak.

## 2.3 LyoCycle

Table 1 Process schedule of the 1. Experiment whose time sequenced FD profile is shown in Figure SS including different phases of the FD product (Lactose 5%)

| Step | Temperature (°C) | Time (minutes) | Cumulative time (min) | Ramp/Hold | Set pressure (μbar) |
|---|---|---|---|---|---|
| Equilibrium | 20 | 30 | 30 | H | - |
| Freezing | -45 | 130 (0.5°C/min) | 160 | R | - |
|  | -45 | 180 | 340 | H | - |
| Re-heating | -32 | 65 (0.2°C/min) | 405 | R | - |
|  | -32 | 180 | 585 | H | - |
| Re-freezing | -45 | 65 (0.2°C/min) | 650 | R | - |
|  | -45 | 180 | 830 | H | - |
| Re-heating | -32 | 65 (0.2°C/min) | 895 | R | - |
| Re-freezing | -40 | 40 (0.2°C/min) | 935 | R | - |
| Freezing hold | -40 | 60 | 995 | H |  |
| Primary drying | -40 | 180 | 1175 | H | 300 |
|  | -25 | 75 (0.2°C/min) | 1250 | R | 300 |
|  | -25 | 4000 | 5250 | H | 300 |
| Secondary drying | 20 | 225(0.2°C/min) | 5475 | R | 300 |
|  | 20 | 600 | 6075 | H | 300 |

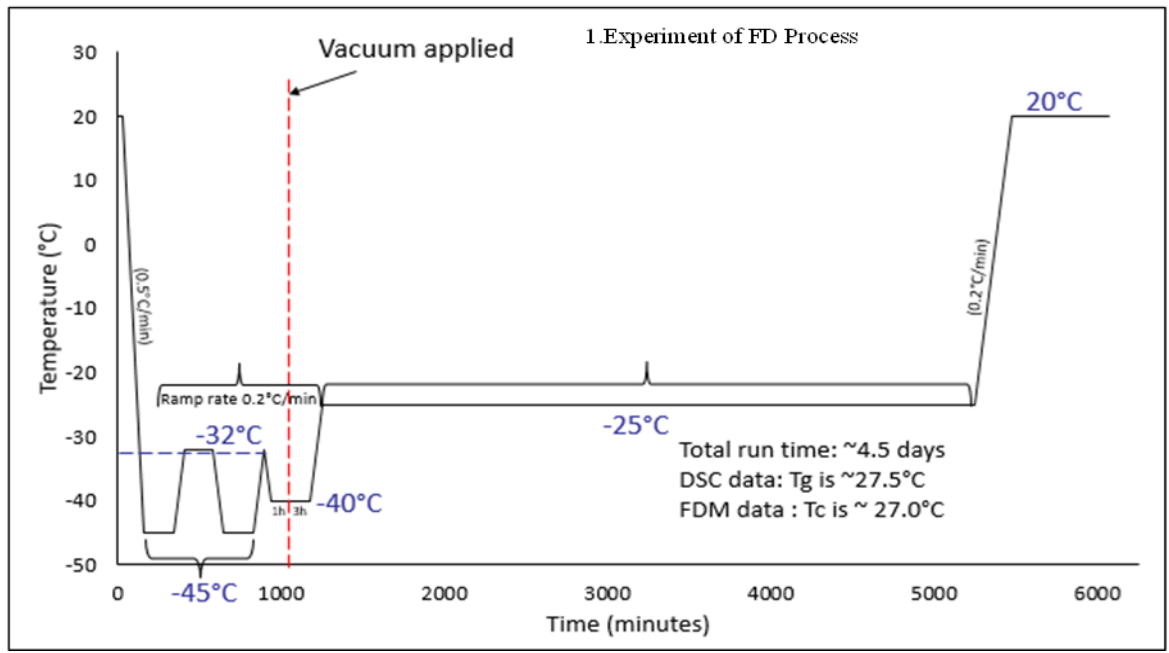

Figure 1. As displayed in Table 1, FD process of the product (5% lactose) in ice form which has been observed and analysed by TVIS. (1. Experiment)

Table 2 Process schedule of the 2. Experiment whose time sequenced FD profile is shown in Figure SA including different phases of the FD product (Lactose 5%)

| Step | Temperature (°C) | Time (minutes) | Cumulative time (min) | Ramp/Hold | Set pressure (µbar) |
|---|---|---|---|---|---|
| Equilibrium | 20 | 30 | 30 | H | - |
| Freezing | -45 | 130 (0.5°C/min) | 160 | R | - |
| | -45 | 180 | 340 | H | - |
| Re-heating | -32 | 65 (0.2°C/min) | 405 | R | - |
| | -32 | 180 | 585 | H | - |
| Re-freezing | -45 | 65 (0.2°C/min) | 650 | R | - |
| | -45 | 180 | 830 | H | - |
| Re-heating | -32 | 65 (0.2°C/min) | 895 | R | - |
| Re-freezing | -40 | 40 (0.2°C/min) | 935 | R | - |
| Freezing hold | -40 | 60 | 995 | H | |
| Primary drying | -40 | 4500 | 5495 | H | 300 |
| Secondary drying | 20 | 300(0.2°C/min) | 5795 | R | 300 |
| | 20 | 600 | 6395 | H | 300 |

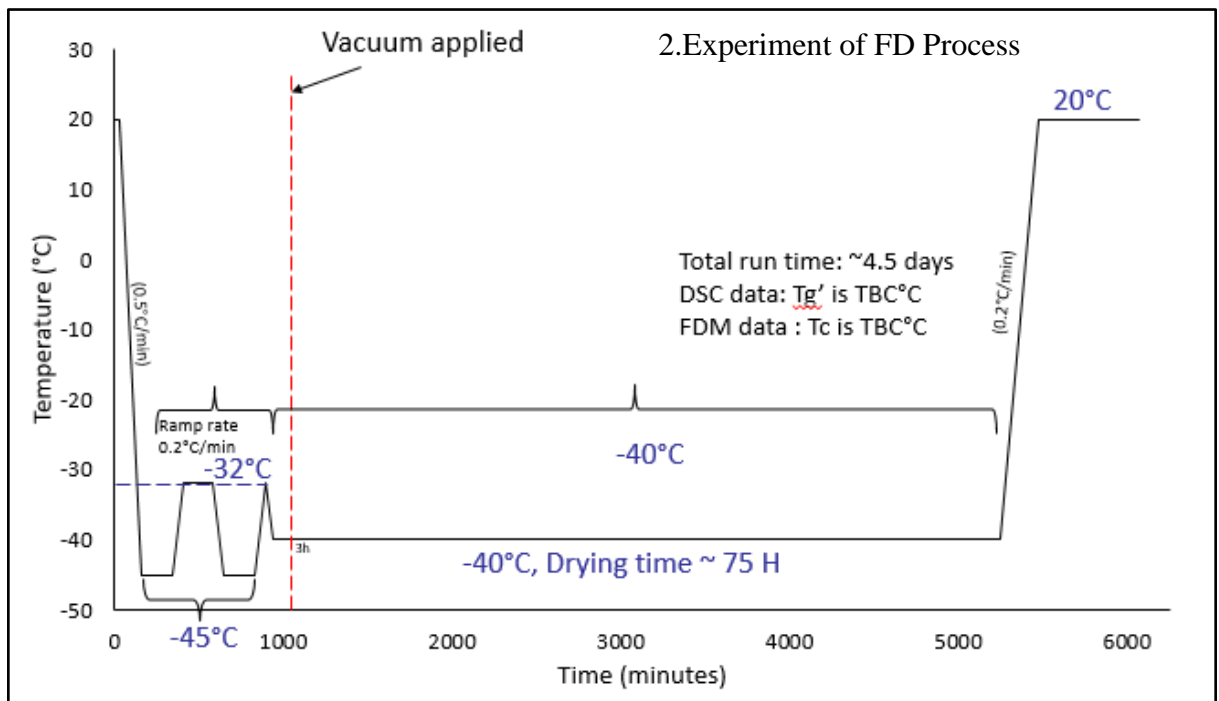

Figure 2. As displayed in Table 2, FD process of the product (5% lactose) in ice form which has been observed and analysed by TVIS (2. Experiment)

## 2.4 Sample layout of the vials

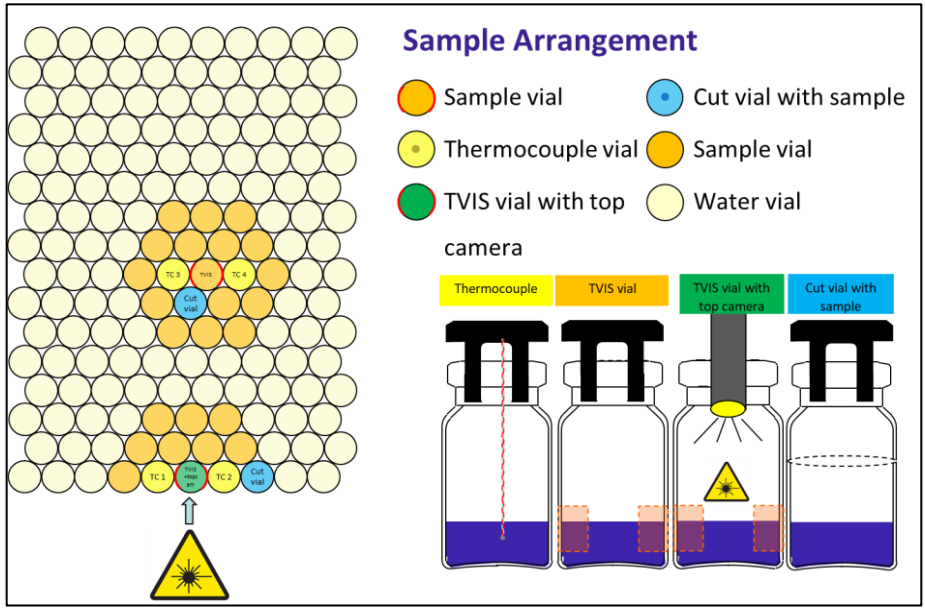

Figure 3.  Sample layout of the test vials

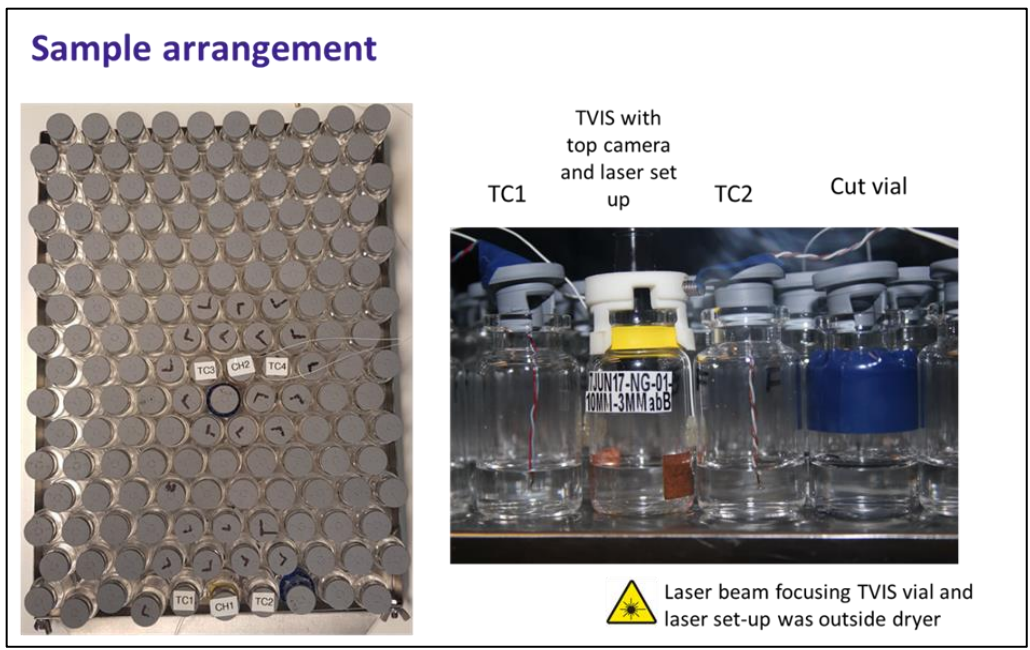

Figure 4. Sample arrangement in FD chamber

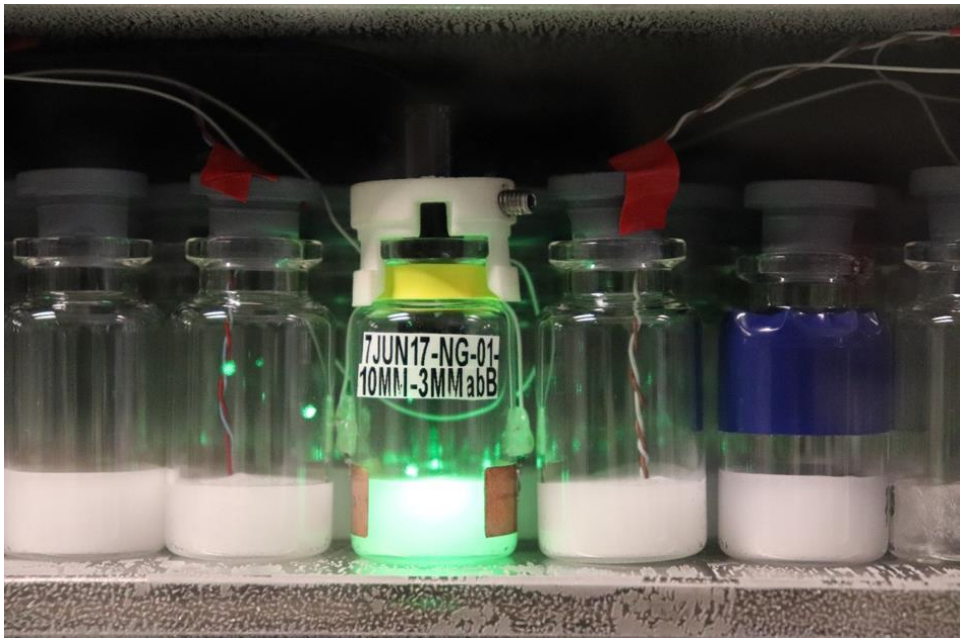

Figure 5. Test Samples in the FD chamber

**2.5 System hardware and configuration**

The hardware utilized within the proposed system consists of a high resolution miniature (CMOS) cameras with an image pixel resolution of 10 micron, and with the focal length of $f = 14$ mm. for video (.avi) image recording and camera software utility (Figure 6) . The Image data collection process in the freeze-dryer chamber is a challenging issue, where the operation temperature, high humidity level and vacuum level are the important factors as the camera has to resist against such a very low (-40º) temperature, humidity and vacuum conditions. In the experiments the camera video recording software utility is used to record 71 hour lasting FD process which corresponds to 1.2 TB data volume even after its data compression. Laser light illumination of FD product surface is done either by side-view technique (through FD Plexiglass™ window) or fiber-optics utility via FD chamber pass-through mechanism.

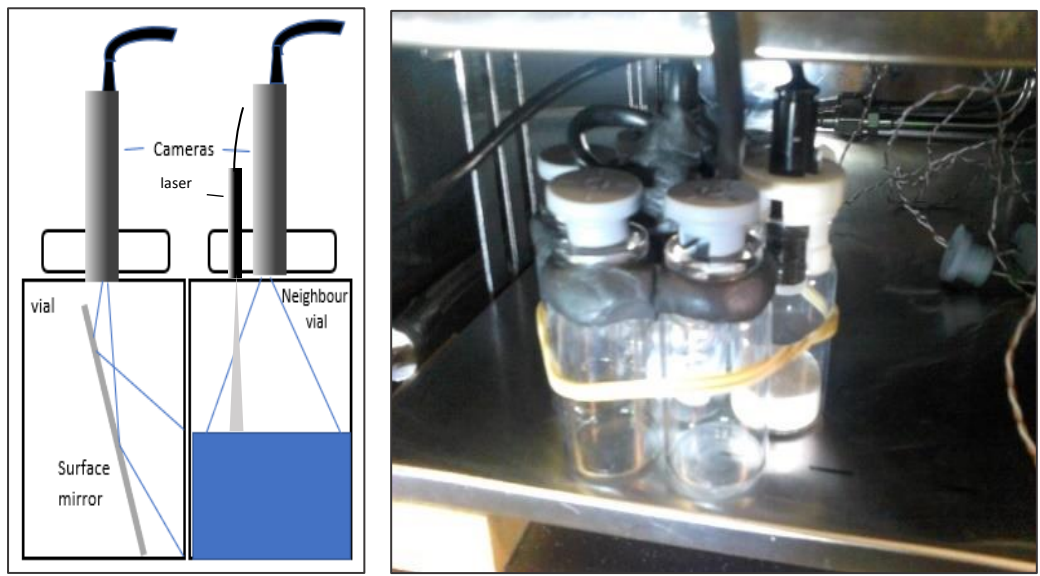

Figure 6. Miniature camera-in-vial configurations (on the left) in which one camera is used for side view of product's volumetric change (optional) and the other camera to observe product surface along the FD process stages and also performs video image data collection during the entire process with laser light illumination via fiber-optics that passed-through into the FD chamber (on the right). The different vial configurations include : thermo coupled vial , TVIS vial, camera-in-vial.

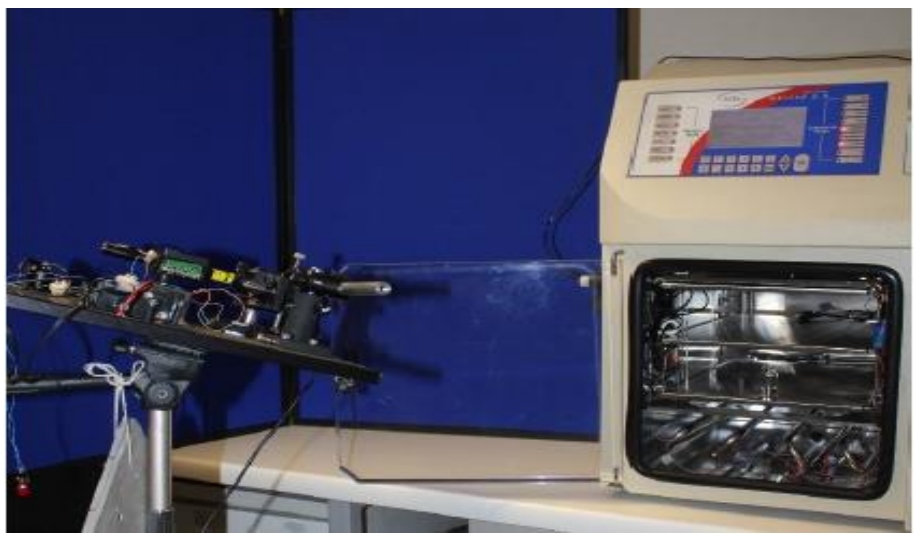

Figure 7. FD system with external laser illumination facility (as an alternative to fiber optics) via FD chamber window. The laser light flash is activated in every 1.2 minutes during video recording by camera-in-vial system.

As is shown in Figure 6, side-view and top-view camera configurations observe different product structural characteristics at different FD stages. In terms of non-invasiveness, the Laser light energy (<1mW) and its exposure (1-2 seconds) are too small to effect product's conditions. As the system is highly flexible, it would be possible to change any of its parameters or component keeping the system in same form, like laser wavelength, camera types, laser energy and its illumination time, texture measures and AI classifier.

The ultimate proposed system configuration is shown in Figure  and system operation in Figure 9 with the continuous real-time process loop for micro-collapse stage (and also $T_c$) detection with its primary components. The system function includes automated image sampling, texture analysis, AI based system training and product states classification process respectively. The training process of the system is expected to be done with adequate data with a minimum inconsistency and also optimized system parameter options like data discretization method, network construction threshold, ROC and  multi-net options, etc. for ultimate system optimisation.  Further to system optimisation, in case of a need to increase the classifier accuracy, a multi-classifier system (MCS) would be used in which at least 3 separate classifiers with different configurations and parameter options are merged whose outputs yield one single results by voting. Such MSC systems had earlier used by several studies [1][18][19]

### 2.6 Test material used in the FD experiments

D-Lactose monohydrate was purchased from Fluka (UK) and used, as supplied, in the preparation of 5% w/v lactose solution in ultrapure water. 3.4 g aliquots of 5% w/v lactose solution to cluster of twenty-seven 10 ml type 1 tubular clear glass vial (Schott, Hungary), to two TVIS measurement vials (Figure 5) and two cut 10 ml glass vials. Remaining glass vial filled with 3.4 g ultrapure water. The test materials' vial arrangement is shown in Figures 3 and 4.

### 2.7 Design of  FD experiments and their process scheduling

Two sorts of complementary FD experiments have been designed to gather various parametric system data along the process including TVIS and LSI methods whose main parameters are shown in Table 1 and 2.  We have demonstrated two FD experiments with sublimation one above and one below Tg' of lactose formulation. The first experiment targets a clear distinction between the two states of the FD product like non-collapse  and micro-collapse forms. Where the corresponding FD time gaps are 17.5-19h   and   20.5-22.5h respectively. The 2. Experiment rather focuses on different forms of dry layer whose thickness varies in regards to FD parameters option and in this case the main goal is  to test different dry layer thickness to develop a "dry layer volume invariant" method by avoiding any possibility of  dry layer volumetric correlation effect on micro-collapse state identification by LSI method.  By this approach, LSI method (or Intelligent LSI  by the inclusion of AI classifiers) would be capable of detecting a micro collapse state of FD product against any volumetric effect (thickness) of dry layer state. In both experiments we have used two annealing steps so we can use 2. annealing data for temperature calibration. The temperature calibration data help to calculate Ĉ″peak. Overall system operation is shown in Figure 9 and 10.

## 3. Results and discussion

### 3.1 Laser speckle imaging with a fibre-optic laser illumination

In principle, the laser speckle imaging (LSI) phenomenon may be described in terms of a coherent light (e.g. laser) and rough surface interaction in the scope of light physics and also described by its statistical property of scattered light distribution over the imaging domain. At very basic, when optically rough surface is illuminated by a coherent light, then the scattered light exhibits a particular intensity distribution making the rough surface to be covered by granular structures called speckle effects [5].

Laser speckle imaging have many application areas from medical imaging [8] to pharmaceuticals [1] and have been also used in industrial material identification [6][7]. The laser speckle image formation and its analysis at the observation domain by use of a digital camera (e.g. CCD or CMOS) is a multi-parametric task and its image's statistical property heavily depends on its system geometry (e.g. laser illumination angle). In the image formation domain a basic light amplitude at point A (e.g. at a single pixel of CCD camera matrix) may be formulized as [5];

$$I_j(A) = |I_j| e^{i\varphi j} \quad (1)$$

The complex amplitude $I_{com}$ in the image domain at point A would be written as;

$$I_{com}(A) = \frac{1}{\sqrt{N}} \sum_{j=1}^{N} |I_j| e^{i\varphi j} \quad (2)$$

Where;
$I_j$: basic light amplitude of a surface element $j$
$A$: Point on image domain (single pixel of CCD matrix)
$\varphi_j$: Random phases of light at $j^{th}$ surface element.
$i$ : Imaginary part
$N$ : total surface elements

In the image domain the speckle pattern is called "the contrast" (C) which may be stated by the Formula 3 as ;

$$C = \frac{\sigma}{\langle I_{mean} \rangle} \quad (3)$$

Where;

$\sigma$ : standard deviation of polarized speckle pattern (standard deviation of spatial intensity variations)
$I_{mean}$ : Mean of intensity (spatial average)
In the formula 3, C is speckle contrast (0 < C < 1) whose ideal value is 1 for a fully developed contrast image.
By a fibre-optic technology a laser light transmission is possible by which a laser light at specific wavelength and intensity would be conveyed into the freeze dryer chamber via vacuum resistant pass-through connector. Such utility is used for product (e.g. lactose) surface illumination by a low energy (<1mW) laser beam at λ=532 nm wavelength at a specific angle (e.g. 5° angle with the product surface normal) to generate laser speckle effect on the product surface. A video image of the illuminated surface is then taken by a camera-in-vial (MiniCam™) during the whole FD process period.

### 3.2 LSI image sampling technique

LSI image samples are selected in highly speckled areas on the global LSI image of FD product surface illuminated by laser light at λ=532nm and whose size is *1257x944* pixels. The image samples have to be at optimum size (e.g. *50x50, 80x80* pixel windows) as each sample should contain enough texture primitive (repetitive pattern) and the texture pattern also has to be homogenously distributed over the sampling area for max accuracy of textural calculations (Equations 4 – 8). The system specification for LSI application is shown in Figure 6 and 7. The image samples are also shown in Figure 11.

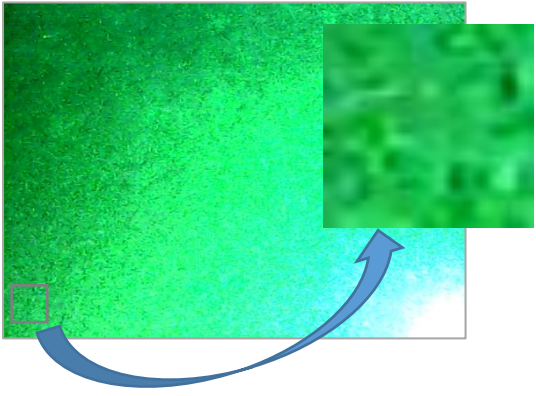

Figure 8. Sampling process of a LSI image with 50x50 pixel window Containing the textural characteristics of the laser illuminated FD product surface

### 3.3 Laser speckle image texture analysis algorithm

If a texture primitive is repeated in an image in a systematic two dimensional (*x,y*) form, it is called a textural image and it may be possible to quantise such textural image to calculate a single value by use of different texture measures (Table 3) where each one proceeds a specific pixel based arithmetic. Some of the texture operator examples are already formulized [10] between the Equations 4 and 8. The whole image sampling area (e.g. 50x50 pixels) is scanned by the texture algorithm and the final cumulative texture value is calculated (Figure 8)

$$Variance_{Russ} = \sqrt{\Sigma(centralpixel - neighbor)^2} \quad (Russ) \tag{4}$$

$$Variance_{Levine} = \frac{1}{area}\Sigma(Central\ pixel - mean)^2 \quad (Levine) \tag{5}$$

$$\sigma = \sqrt{Variance_{Levine}} \quad (Sigma) \tag{6}$$

$$Skewness = \frac{1}{\sigma^3}\frac{1}{area}\Sigma(Centralpixel - mean)^3 \quad (Skewness) \tag{7}$$

$$Std.Deviation = \sqrt{\frac{\Sigma(x - x')^2}{n}} \quad (St.deviation) \tag{8}$$

Each texture measure [10] has different characteristics and function like scale-invariant, rotation-invariant, pixel noise-tolerant, etc. and treated as an independent attribute in an AI system (e.g. Bayesian network, k-nn, SVM) data set.

### 3.4 Artificial Intelligence classifiers (Bayesian networks and k-nn algorithm)

Two classifiers (Bayesian networks and *k-nn* algorithm) have been used for speckle image textural data analysis to distinguish between the "normal" and "micro-collapse" stages of the Lactose product. Data attribute for each texture measure are shown in Table 3 for each FD stages. Within the Bayesian classifier (PowerPredictor™) the options and parameters are use as follows : equal frequency discretisation method, equal training/test sets and minimum network threshold (*t = 0.1*) for max number of network attribute connection. Whereas in *k-nn* classifier algorithm "standardized" parameter option was used by which the classifier utility centres and scales each column of the training data by the column mean and standard deviation respectively.

**Bayesian Network (BN) :** It is very commonly used AI method for data analysis., BN method is well known as dealing with the bad data with high uncertainties in high efficiency. Its operational principles may be described by its generic formula 9.

$$P(X) = \Pi_i P(X_i pa(X_i)) \tag{9}$$

where : P(X) is the joint probability distribution which is the product of all conditional probabilities. Pa($X_i$) is the parent set of $X_i$ (e.g. class node to decide normal/micro-collapse product) For the experiments the BN utility called PowerPredictor$^{TM}$ is used (Figure 19)

**K-NN (k-nearest neighbour) classifier algorithm :** It is a non-parametric method commonly used for classification, where its input includes the "*k*" closest training examples in the feature (observation) space. In the classification process the object is classified by a plurality vote of its neighbours [11]. In the classifier, the point "*x*" that represents the object to be classified in the feature space is assigned to the class of its closest neighbour in the feature space, shown by the formula 10.

$$C_n^{1nn}(x) = Y_{(1)} \tag{10}$$

**Y** is the classes label of X., The index (1) of Y indicates the k=1 nearest neighbour classifier type. *k* is user defined constant and **C** is the feature space of observations.

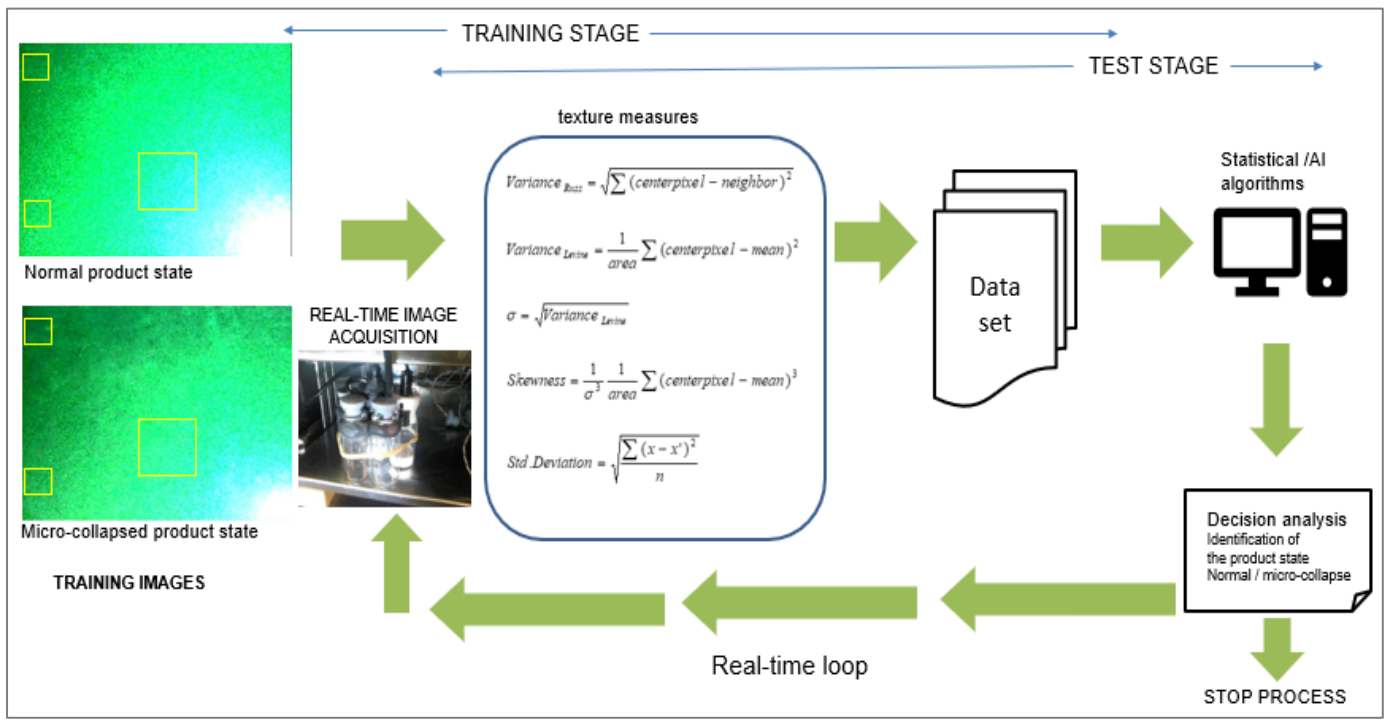

Figure 9. The proposed system configuration with the real-time process loop for micro-collapse stage (and also T$_c$) detection is shown with its primary components. Each component is flexible enough to separately upgradable with another alternative instrumentation and algorithm (e.g. camera type, laser wavelength/power, classifier algorithm, etc.)

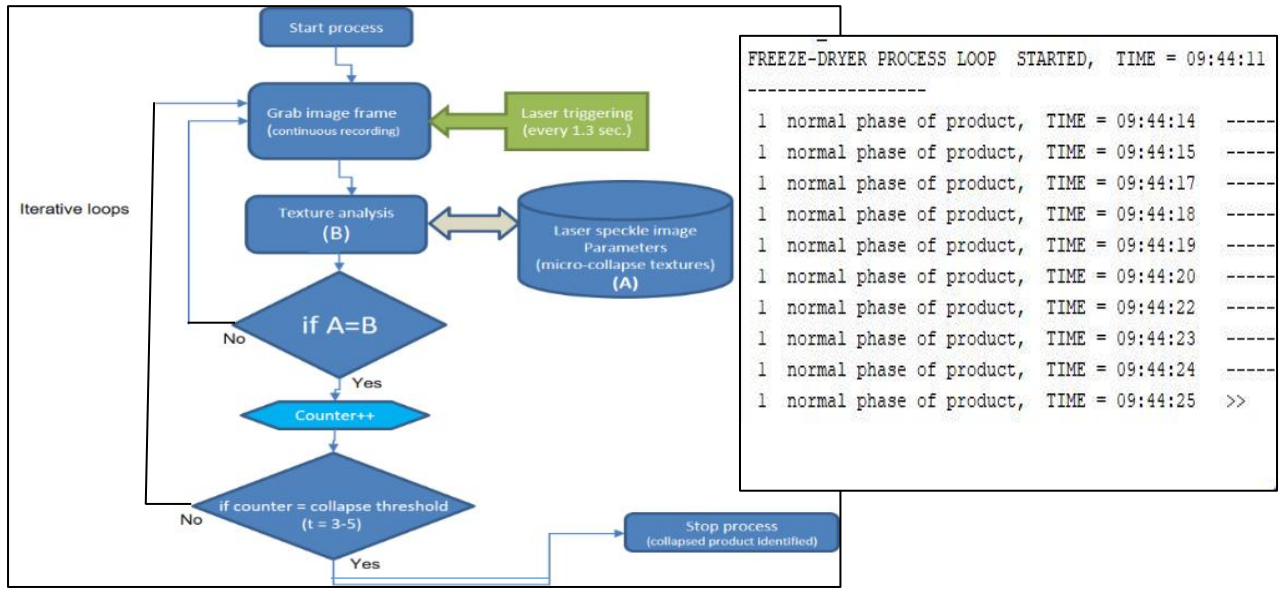

Figure 10. Flow chart of the system to operate in real-time (on the left) as shown in Figure 9 and the system output as is in real-time simulated operation by the product observation loop.

# 4. Experiments

**4.1 First experiment to identify micro-collapse and product frozen states** The time schedule of 1.experiment and its process parameters profiles (temperature, vacuum, etc.) can be seen in Figure 1. This period refers to the initial stage of FD process where the product (5% lactose) was in ice form and its has been observed and its parameters and data stream were recorded by both TVIS and Laser Speckle Imaging (LSI). The experimental test characterisics are shown in Figures 12, 13 and 15.

**4.2 Second experiment to generate/observe different volume of dry layers to investigate their micro-collapse state correlation**
The time schedule of 2.experiment and its process parameters profiles (temperature, vacuum, etc.) can also be seen in Figure 2. The formation of product's different drylayers have been observed and their parameters and data stream were recorded by both TVIS and Laser Speckle Imaging (LSI) technique. The experimantal cahracteristics are shown in Figure 14.

**4.3 Application of Intelligent laser speckle Imaging *(ILSI)* technique** The proposed *ILSI* technique demonstrates the chain of processes in the following order; **a)** sequential video image data acquisition of a periodically laser-illuminated (e.g. every 1-2 minute) product surface **b)** laser speckle image data extraction from the video sequences, **c)** time location analysis for before and after micro-collapse product stages in the video sequence for data extraction, **d)** data analysis and classification to determine micro-collapse product stage and $T_c$, **e)** Scanning Electron Microscopy (SEM) analysis after the FD session to verify product micro-collapse. In the experiments, the entire FD process has been recorded by video recorder utility software MiniCam[TM] in avi format for about 71 hours, then the laser speckle images of 60 cases are extracted from the entire 71 hour video image data by visual inspection with their class-categorisation as "normal/micro-collapse" (figure 11) using parametric calculation (figure 12), and based on the image sampling areas A, B and C of different laser diffuse reflection bands (Figure 18), and then quantised by the texture measures described by the formulas 4-8. These whole chain of process was to construct a data set with 60 cases and 2 classes (labelled as normal and micro-collapse product phases) including 27 attributes (matrix of *60x27* as some part of it is shown in Table 1). The data set contains 9 texture measures for 3 different laser light diffusion bands. Following the classification processes by BN and *k-nn*, the classification results particularly obtained by Bayesian Network classifier (PowerPredictor[TM]) where the texture measure Skewness(*3x3*) automatically selected and applied on sampling area C on image, was playing a primary classification role (Figure 18). BN classification and k-nn classifier, both yielded 100% classification accuracy as exhibited by Figure 19. In BN classifier (PowerPredictor[TM]) equal frequency discretisation method was used instead of equal width or entropy based options, Meanwhile equal size of training/test sets were also used with the minimum network threshold (*t = 0.1*) for max number of network attribute connection. In *k-nn* classifier algorithm there were less option for its basic level usage such as ;"standardized" parameter option was used by which the classifier utility centres and scales each column of the training data by the column mean and standard deviation respectively. As seen in Figure ss, the normal product stage before the micro-collapse in primary drying period lasted about 1.5 hours which then ended up with product's micro-collapse period lasting 2 hours exhibiting very sharp C"$_{peak}$/P$_F$ turning point at about 21.hour. These two periods characterised the product structure as clearly shown in Figure ww. With the specifically selected wavelength (λ=530nm), the laser illumination of the product surface helped increase the clear visibility of structural product phase characteristics by generating the laser speckle effects on its surface. The textural structures of laser speckle images for normal and micro-collapsed (Figure 11) product states are not always distinguishable by a naked eye. Hence needs to be quantised by the texture algorithms as already explained in Chapter 4.2.

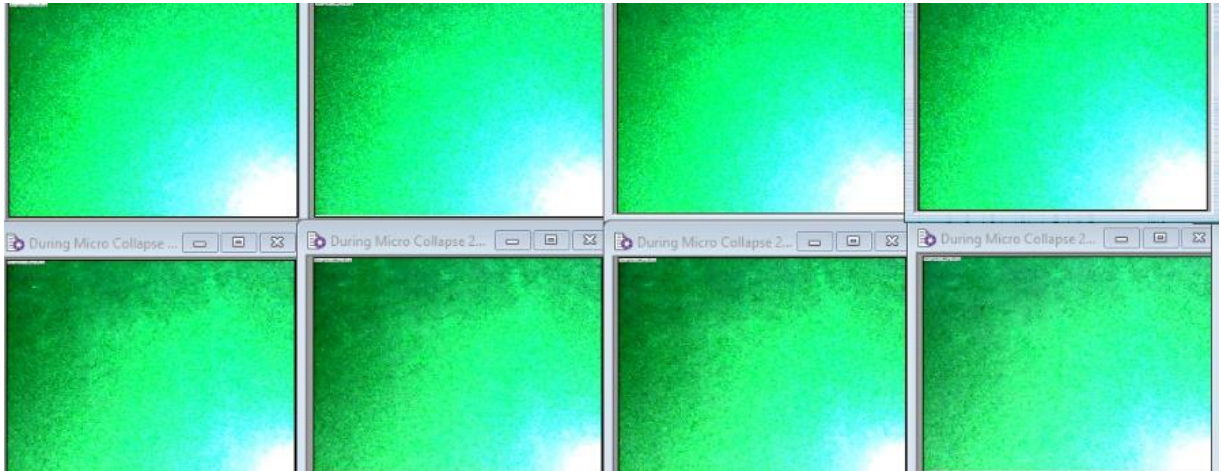

Figure 11. Even though a laser illumination of the product surface definitely increases the visibility of structural Product phase characteristics by generating the laser speckle effects on its surface, the textural structures of laser speckle images for normal (upper row) and micro-collapsed (lower row) product states are not always distinguishable by a naked eye. Hence needs to be quantised by the texture analysis and then classified by AI statistical methods for clear automated identification in a continuous real-time FD process.

As far as such texture analysis is concerned, large window size produce large edge effect at the class edges but provide more stable texture measures than small windows. In return, small window size is less stable but has smaller edge effect (22).

### 4.4 Application of TVIS (Through Vial Impedance Spectroscopy) to FD process

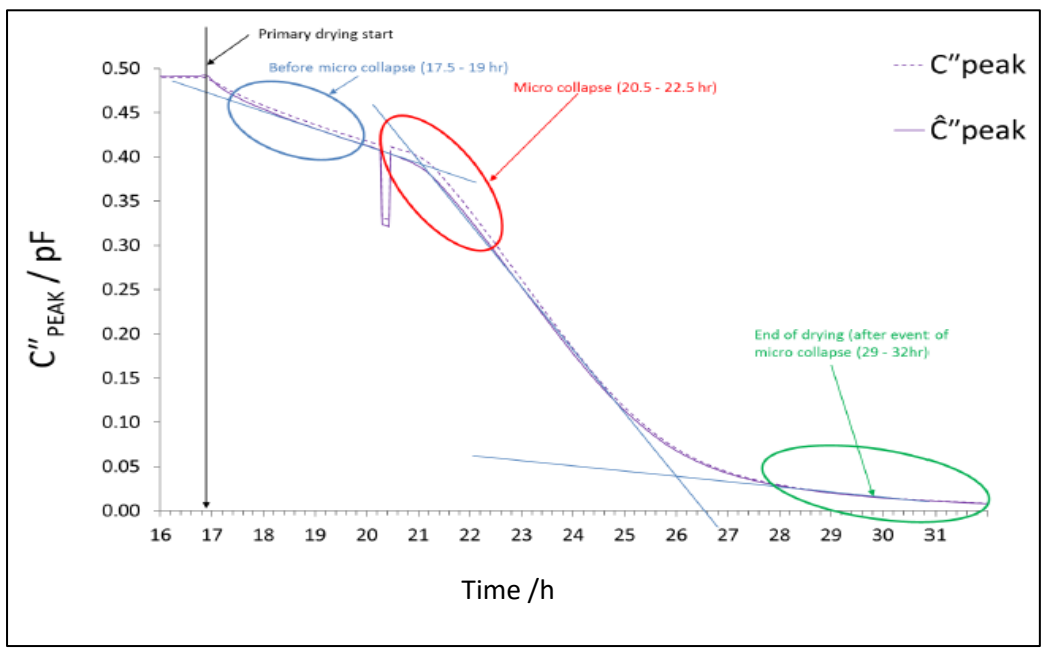

Figure 12. 1. Experiment: In regards to video image recording of FD process, approximate micro-collapse time gap is to appear between 20.5 – 22.5 hours. In the graph C$^{//}$ peak is the imaginary Part of capacitance (ice mass & temperature), and C$^{\wedge //}$ peak is imaginary part of capacitance (standardized by temperature C$^o$)

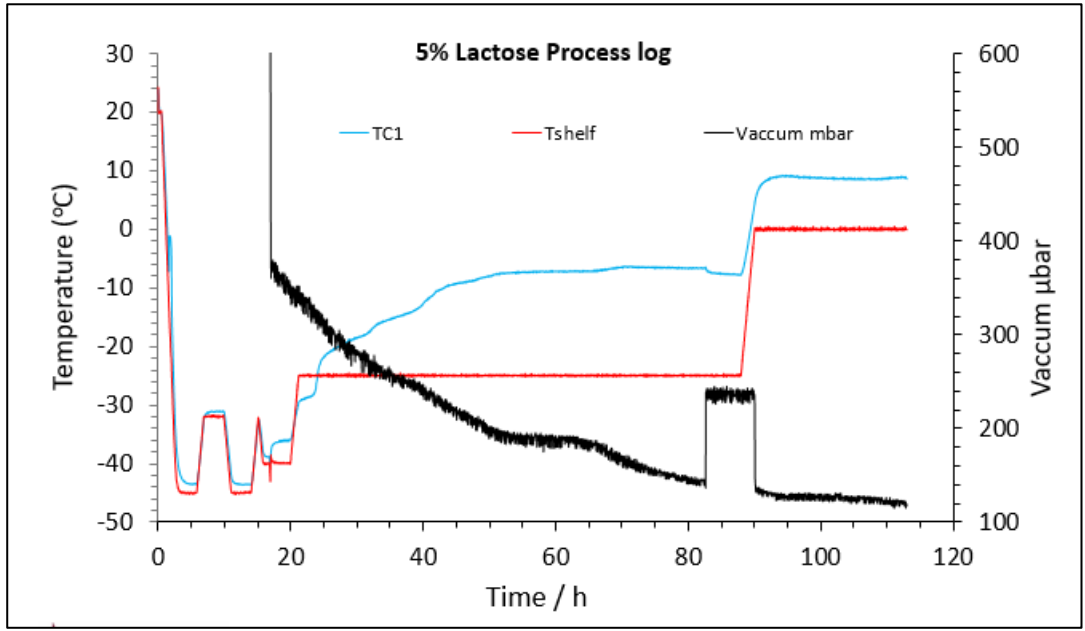

Figure 13. first Experiment's characteristics for FD Process of 5% Lactose

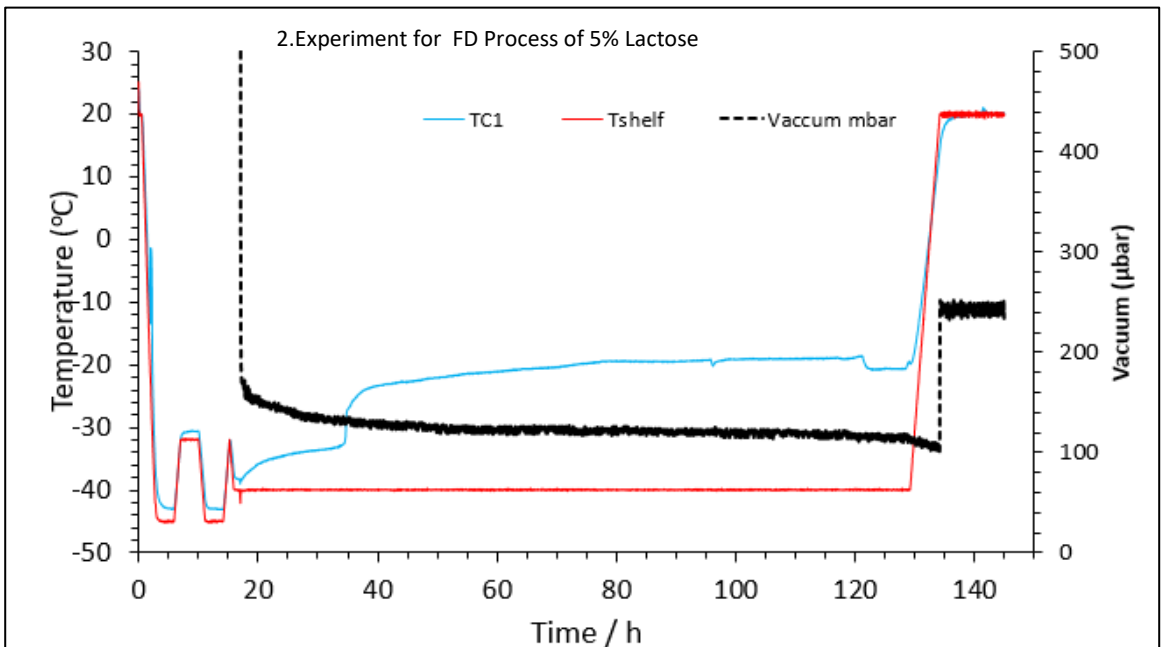

Figure 14. Second Experiment for FD Process of 5% Lactose

**4.5 LSI image data extraction for the different product states for their analysis**

For the laser speckle image data extraction from the video image sequence, the whole time sequence is transformed into a Polynomial graph (solid sinusoidal curve) which helps locate the before and after product micro collapse state regions whose centres are marked by vertical dash line in Figure 15. At its initial development stage, the model refers to TVIS method for definition of each product state's specific time range in the FD process. In Figure 15, the centres of before and after micro-collapse state regions are marked by solid vertical lines. The Figure 15 is also comparable with Figure 12 as the first one configured by TVIS method calculations.

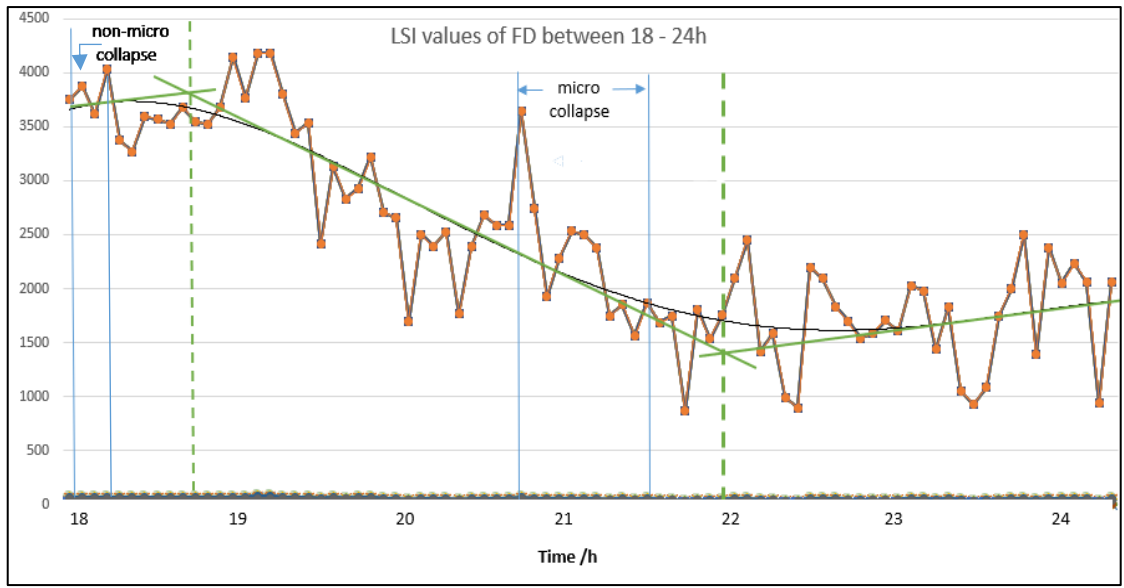

Figure 15. First Experiment: Polynomial model of the Laser speckle (LS) values of most discriminant texture measure (levine), covered by 6 hour FD process period helps locate the non- micro-collapse and micro collapse states regions which are clearly visible to naked eye, but some portions of graph which are highly collated to micro collapse stage can be analysed by Bayesian classifier to distinguish both categories. As is seen only texture measure Levine2 (*5x5*pixel kernel) exhibit high level of Laser Speckle features correspond to product states class variations.

### 4.6 I. and II. Experiments to generate different volume of dry layers to investigate the micro-collapse state correlation

First and second LSI experiments have been achieved to prove the advantage of ILSI (Intelligent LSI) technique over TVIS Method which can differentiate the different layer states those sharing the same capaciatance value (Figure 16). In the FD experiments different thickness of dry layers production was targeted (Figure 17). The process was designed for a further verification of discrimination capability of the proposed system by which micro-collapse product state would be distinguished from the different dry layer thickness of the product. This was to avoid any possible correlation between the micro-collapse state and volumetric changes of dry layer of the product.

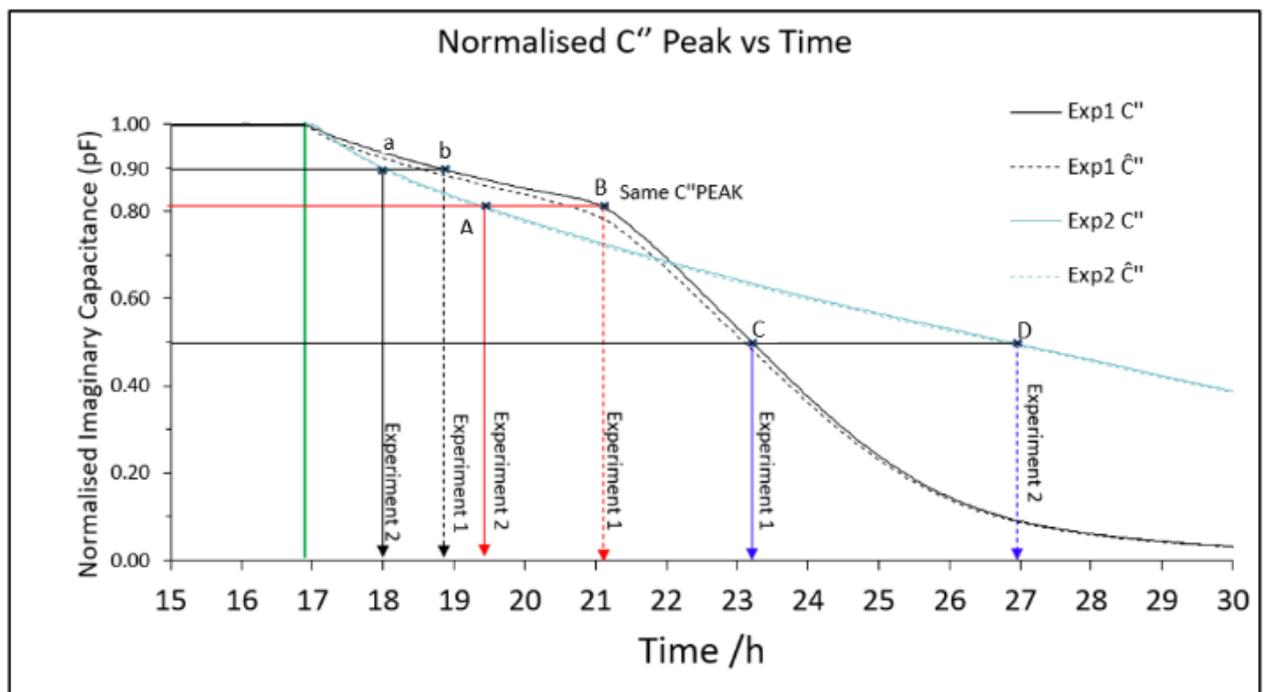

Figure 16. Normalised imaginary capacitance vs time

F<sub>PEAK</sub> calibration curve

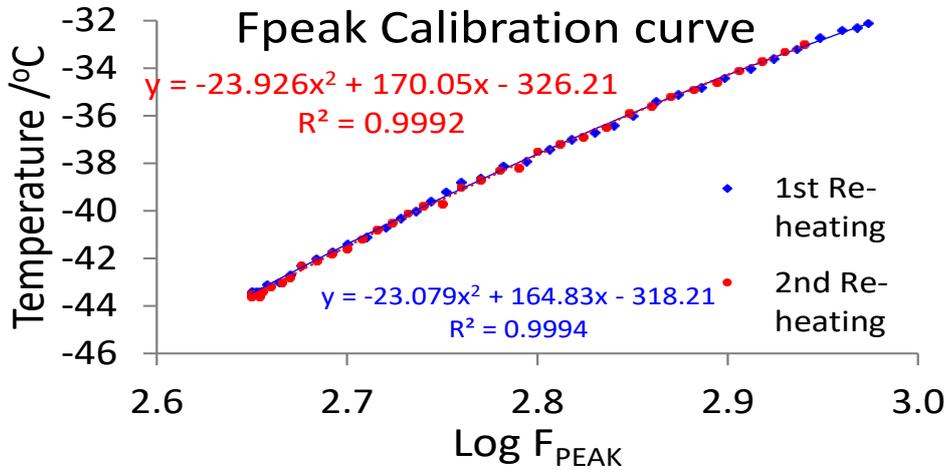

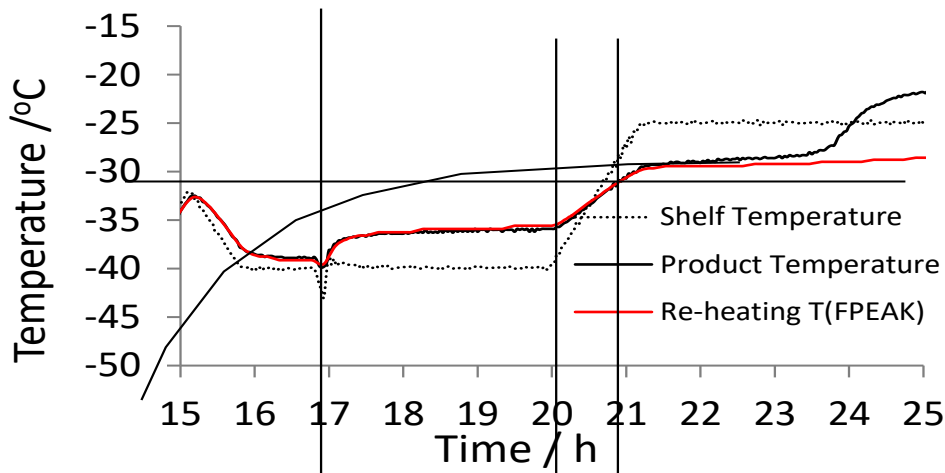

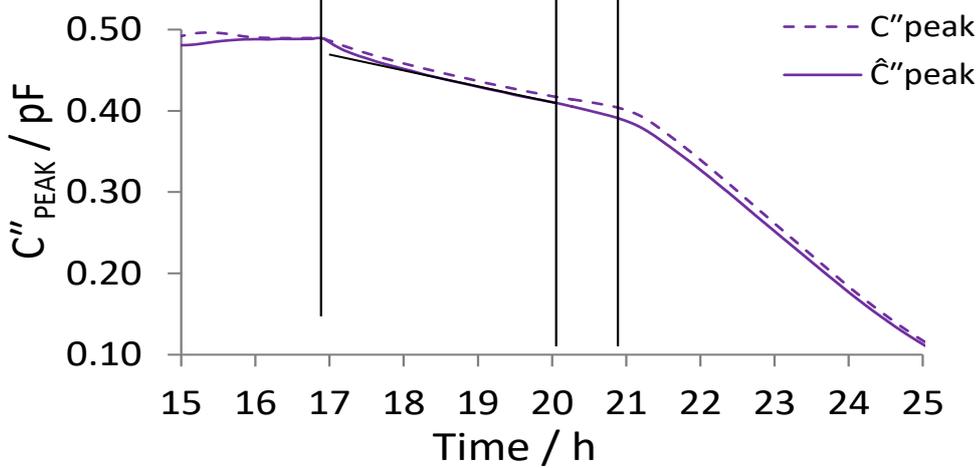

**Figure 17.** First and Second LSI experiments are exploited to prove the advantage of ILSI (Intelligent LSI) technique over TVIS to discriminate highly identical product states. The dry layer states A and B at different thickness whose LSI values collected in 1. And 2. Experiments have been classified with 91% accuarcy even though both have same PF value (0.80). Meanwhile drylayer D and Micro-collapse state C are also to be classified via Bayesian classifier as both have common PF value (0.50)

The classification accuracy of 91% for the two different FD product states (dry layer (A) and micro-collapse state (B) has been obtained. The classification results and the process profile graph are shown in Table V and Figure 17 respectively.

## 4. 7 Classification by Bayesian Networks and *k-nn* algorithm

Exploitation AI technique is almost inevitable to process particularly the inconvenient image data sets with some degree of uncertainty. On this basis, AI based classification methods [1][20] have also an extensive capacity of unveiling invisible information to distinguish very identical surface features. As every classifier algorithms have different statistical approach and principles, within this work two complementary classification methods *(Bayesian and k-nn)* are used following the texture analysis, whose results are presented in Table 4 and 5.

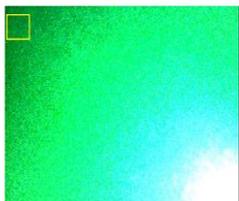

Table III. Sample Data set of laser speckle image texture results for "normal" and "micro-collapse" FD stages. The image segments (shown at bottom-right) are quantised by the different texture measures by Formulas 4 -8 (Chapter 4.2). The 3x3 and 5x5 values shown in the attribute labels indicates pixel window kernel size of image scanning operator to scan entire image. The data table represents only some part of (1/3) the whole data set used for the classification (highlighted attribute columns indicate the relatively high variations between the two classes normal and micro-collapse)

Sampling selection process on each LSI is usually made manually by naked eye and in normal imaging conditions can not be done by an automated process due to the unpredictable displacement of the textural speckle areas in each single image. This is because of unpredictable product surface conditions during image acquisition for different product states. But by use of very specific custom design "automated sampling" algorithms [21] such automated process based on histogram operated "image segmentation" could be exceptionally applied in fully-automated industrial real-time systems.

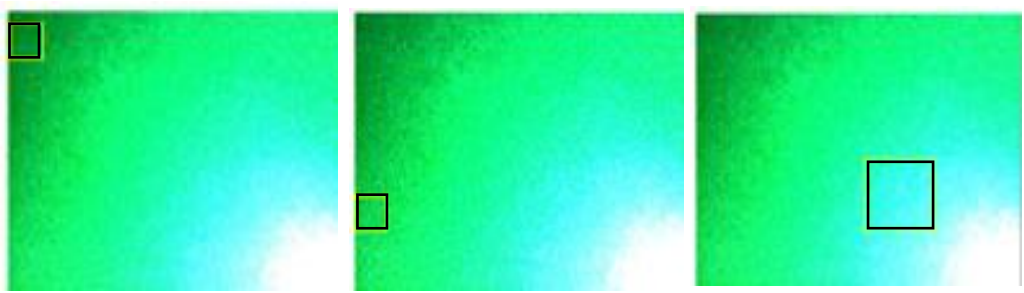

Figure 18. Different sampling areas and sizes with A,B (50x50) and C (80x80) pixel windows on different laser speckle textural bands on the images (on three major laser light diffusion bands other than a central specular reflection area (shown in white)

Table IV. 1. Experiment: classification results obtained by the classifiers with their system optimisation (Figure 19)
(k-NN algorithm is only tested in simulation of real-time operting version of the ILSI system in the first experiment due to its high speed)

| Classifier Type | System parameters | Classification accuracy | Sensitivity | Specificity | Utility name |
|---|---|---|---|---|---|
| Bayesian Networks | t = 0.1, discretisation = equal frequency training/test sets = 1/1 | 100% | 100% | 100% | PowerPredictor™ |
| k-NN algorithm | Standardized training/test sets = 1/1 | 100% | 100% | 100% | k-Nearest Neighbours |
| | | | | | |

Table V. 2. Experiment: classification results obtained for the FD states (Points A and B as is shown in Figure vvv )

| Classifier Type | System parameters | Classification accuracy | Sensitivity | Specificity | Utility name |
|---|---|---|---|---|---|
| Bayesian Networks | t = 0.1, discretisation = equal frequency training/test sets = 1/1 | 91% | 91% | 91% | PowerPredictor™ |
| | | | | | |

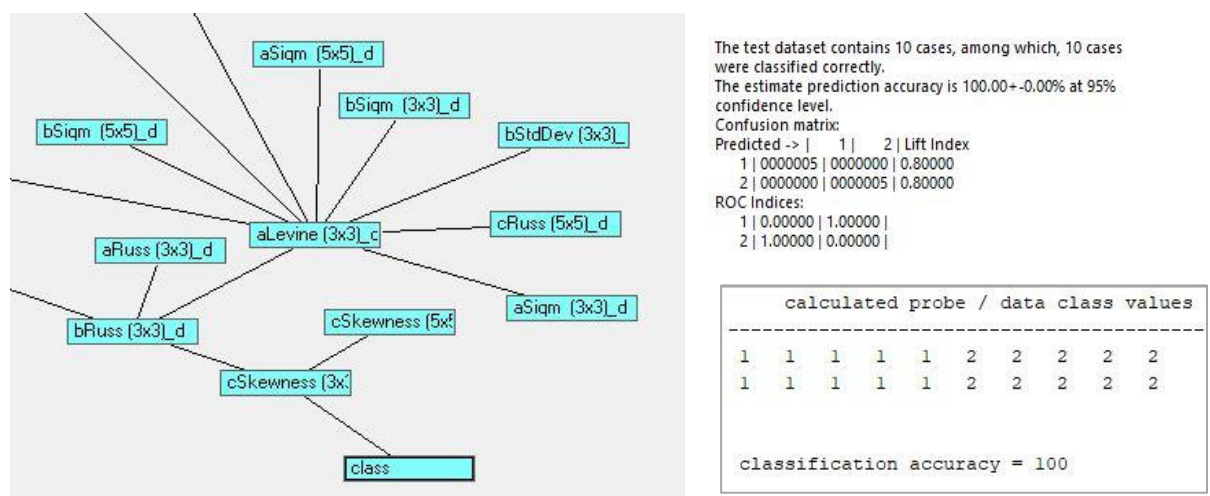

Figure 19. The classification results obtained by automatically constructed BN (on the left) where the texture measure Skewness(3x3) was applied on sampling area C which played primary classification role (Figure 18). Area C was also automatically selected (among the other areas A and B) in data set. BN classification confusion matrix is display on top-right, and screen display of k-nn classifier result is on bottom-right, Both showing 100% classification accuracy.

### 4.8   Scanning Electron Microscopy (SEM) analysis

Scanning Electron Microscopy (*SEM*) [15]  is a supplementary method by which a very close inspection  of the product structure is made after the entire FD process in non-real-time for the absolute confirmation of product micro-collapse state at specific FD time. SEM analysis is inevitable as it justify the all previous laser speckle analysis and classification (stage "a" – "d" as described at the beginning of  Section 4). SEM images of Lactose sample (5%) taken after the FD process displaying the surface  structural characteristics in different product stages like normal phase and micro-collapse as shown in Figure 20. Each scale bar on the images (at the bottom-left corners) is useful tool to compare the size of surface features to the laser wavelength. As a physical principle the laser wavelength has to be shorter than the surface  features to be interacted and it is diffusely reflected back to the camera for LS imaging.

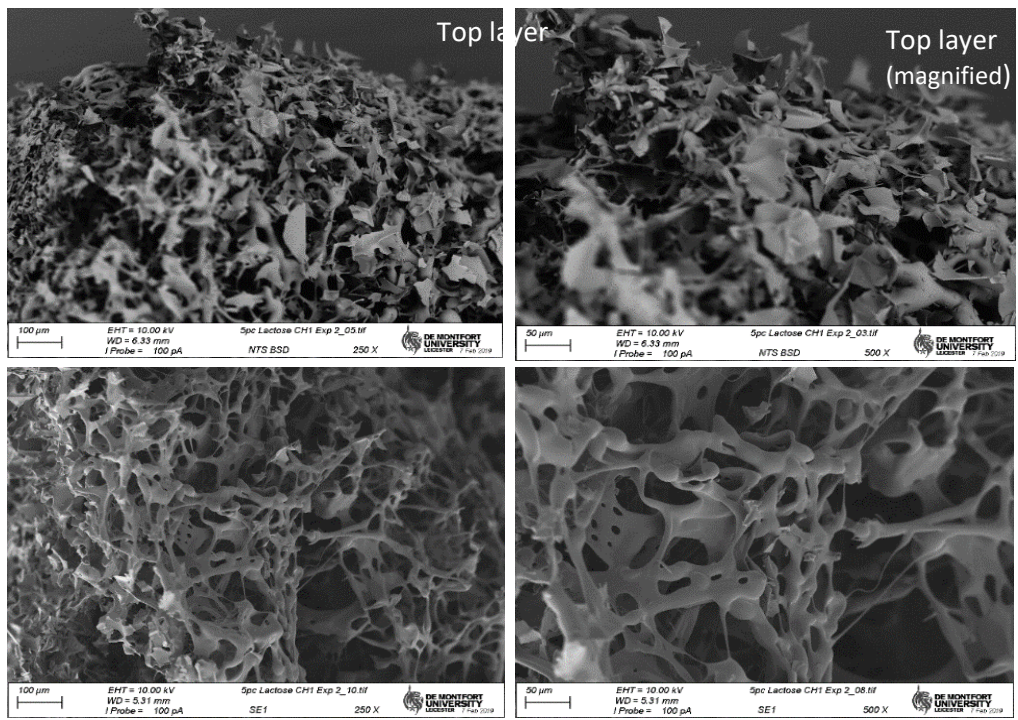

Figure 20 . Experiment 1 .SEM images of the Lactose samples used in the ILSI experiments: normal product states (top row) and micro-collapse states (bottom row) at different scales (250x on the left) and (500x on the right) taken after the FD process displaying their specific surface structural characteristics. The scale-bar at the bottom-left of each images can be comparable with laser light wavelength λ=0.5 µm (530nm) used for speckle imaging, which has to be shorter than the surface features to be interacted with and then diffusely reflected back. The structural product surface differences between the two states can be easily distinguished.

# 5 Conclusion

Camera-in-vial vision system which has been further enhanced in comparison to available classical vision systems by the laser speckle imaging technique and AI classifiers which is called "Intelligent Laser Speckle Imaging" (ILSI) system, leading to a better and more precise product observation during the whole FD process (micro-collapse stage in particular). Within the proposed work, the Laser speckle imaging data were used for specific purpose of micro-collapse state detection which leads to the critical temperature determination. The system real-time operation was basically LSI analysis of micro-collapsed and non-micro collapsed( normal) product states and their classification process which yielded 100% classification accuracy in the 1. And 2. Experiments with its equal sensitivity and specificity.

Within this work a novel approach called "dry layer volume invariant" has been introduced where the thickness of dry layer should not effect the identification of micro-collapse product state in FD process. To achieve this goal, the effects of different dry layers in two different experiments have been investigated and no-effect results have been proven by series of classifications. The other advantage of the ILSI technique over the conventional TVIS sytem would be, the same TVIS output of two different process states can be discriminated by ILSI.

The system instrumentation is a low cost design since it is based on basic off-the-shelf components. The proposed system is open ended to be further developed by simple modifications (e.g. using a suitable laser sources operating in different spectral bands (IR, UV, etc.) and its power selection (at mW levels) , with various options of AI methods (e.g. Bayesian networks, k-nn, SVM, Neural Nets, etc.). The system is capable of operating in real-time by a minor modifications specific to its industrial or scientific application areas and by use of continuous operational loop at high speed which automatically enables to identify product micro-collapse state as fast as possible. As the verification of the system reliability and accuracy were two main factors at high importance, the justification has been made by the results of Scanning Electron Microscopy (SEM) and would be further confirmed by FD microscopy which are the two supplementary methods.